\documentclass[prb,twocolumn,superscriptaddress,
longbibliography,aps,10pt]{revtex4-2}

\usepackage{graphicx}
\usepackage{bm}
\usepackage{color}
\usepackage{epstopdf}
\usepackage{amsmath}
\usepackage{amssymb}
\usepackage{epstopdf}

\usepackage[normalem]{ulem}

\usepackage{xfrac} % better frac

\usepackage{multirow}

\usepackage[urlcolor=blue,colorlinks=true,citecolor=blue,linkcolor=blue,pdfstartview={FitH},bookmarks=false]{hyperref}

\usepackage[table]{xcolor}

\graphicspath{{fig/}{./fig/}{.}}

\begin{document}

\title{Electronic properties and surface states of CeRh$_{2}$As$_{2}$}

\author{Konrad Jerzy Kapcia}
\email[e-mail: ]{konrad.kapcia@amu.edu.pl}
\affiliation{Institute of Spintronics and Quantum Information, Faculty of Physics, Adam Mickiewicz University in Pozna\'n, ul. Uniwersytetu Pozna\'{n}skiego 2, PL-61614 Pozna\'{n}, Poland}
%\affiliation{Center for Free-Electron Laser Science CFEL, Deutsches Elektronen-Synchrotron DESY, Notkestr. 85, 22607, Hamburg, Germany} %nie ma juz tej afiliacji

\author{Andrzej Ptok}
\email[e-mail: ]{aptok@mmj.pl}
\affiliation{Institute of Nuclear Physics, Polish Academy of Sciences, ul. W.E. Radzikowskiego 152, PL-31342 Krak\'{o}w, Poland}

\date{\today}% It is always \today, today,
% but any date may be explicitly specified

\begin{abstract}
Locally noncentrosymmetric CeRh$_{2}$As$_{2}$ exhibits characteristic two-phase unconventional $H$--$T$ superconducting phase diagram. 
The transition from even- to odd-parity superconducting phase is supported by the P4/nmm crystal structure and it can be induced by the external magnetic field.
Dual nature of the Ce $f$ electrons and competition between their itinerant and localized nature can be important for its physical proprieties.
The existing electronic band structure in this material is still under debate. 
Energy scale occurring in the system can be important for adequate theoretical description of its supercomputing properties.
In this paper, we discuss electronic band structure obtained from {\it ab initio} calculations.
We analyze electronic surface states which can be compared with recently obtained angle-resolved photoemission spectroscopy measurement (ARPES). 
Additionally, we present the analyses of the hopping parameters obtained from the electronic band structure.
The As-Rh layers are characterized by much more significant hopping parameters between atoms than these obtained for the Ce-Ce atoms.
This suggests need of a revision of the tight binding model necessary for adequate description of the physical properties of CeRh$_{2}$As$_{2}$.
\end{abstract}

\maketitle

\section{Introduction}

\begin{figure}[!b]
\centering
\includegraphics[width=\linewidth]{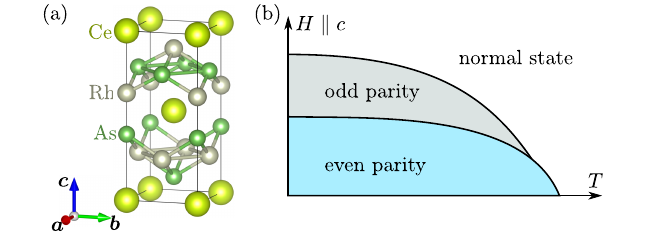}
\caption{
(a) Crystal structure of CeRh$_{2}$As$_{2}$ and (b) its characteristic $H$--$T$ phase diagram for magnetic field along $c$ axis.
\label{fig.schem}
}
\end{figure}

Recently studied CeRh$_{2}$As$_{2}$ exhibits very rare two-phase unconventional superconductivity~\cite{kim.landaeta.21}.
The magnetic field along $c$-axis causes the transition between two different superconducting (SC) phases.
This behavior leads to characteristic $H$--$T$ phase diagram  for ${\bm H} \parallel c$ (see Fig.~\ref{fig.schem}).
Thermodynamic measurements uncover superconductivity below $T_{SC} = 0.37$~K, while the critical magnetic field of its high-field phase is around $14$~T~\cite{kim.landaeta.21}.
The observed SC phases can be tuned by the external pressure, which leads to realization of the second SC dome~\cite{siddiquee.rehfuss.22}.
The $^{75}$As nuclear magnetic resonance (NMR) studies suggests that XY-type antiferromagnetic (AFM) fluctuation occur~\cite{kitagawa.kibune.22}.
The very early theoretical calculations of the ground state support the realization of the AFM order with the magnetic order along $c$ in this material~\cite{ptok.kapcia.21}.
The experimental evidence of the AFM order in CeRh$_{2}$As$_{2}$, with the N\'eel temperature of $T_{N} \approx 0.25$~K, is based on the observation of different broadenings of the nuclear quadrupole resonance (NQR) spectrum at two crystallographically nonequivalent As sites~\cite{kibune.kitagawa.22}.
$T_{N} < T_{SC}$ in ambient pressure and it can suggest a coexistence of AFM and SC orders in CeRh$_{2}$As$_{2}$.
Indeed, the coexistence of AFM and low-field SC phase was reported experimentally~\cite{ogata.kitagawa.23}.
Nevertheless, the relation between $T_{N}$ and $T_{SC}$ is still under debate.
Contrary to this, the is no signatures of the magnetic order in the high-field SC phase~\cite{ogata.kitagawa.23}.

The crystal structure of CeRh$_{2}$As$_{2}$ can be discussed in the context of the Ce atoms, which locally break the inversion symmetry [Fig.~\ref{fig.schem}(a)].
The experimentally investigated $H$--$T$ phase diagram [schematically presented on Fig.~\ref{fig.schem}(b)] exhibits occurrence regions of two SC phases.
This can be associated with the even-to-odd pairing transition~\cite{kim.landaeta.21,landaeta.khanenko.22}.
This finding is supported also by the crystal structure, which possesses a symmetry required to stabilization of the odd-parity SC state~\cite{cavanagh.shishidou.22}.
Theoretically obtained electronic band structure exhibits complex nature with Ce $f$ orbitals in close vicinity of the Fermi level~\cite{ptok.kapcia.21}.
The external magnetic field can lead to the Lifshitz transition~\cite{ptok.kapcia.17}, and, as consequence, to the field-driven odd-parity SC phase~\cite{cavanagh.shishidou.22}.
Additionally, the Ce $f$ electrons and competition between their itinerant and localized part can be crucial for better exploration of the realized phase transitions between two superconducting states~\cite{machida.22}.
The strong correlation realized in the system can be a source of the observed Kondo physics~\cite{cavanagh.shishidou.22,hafner.khanenko.22,christovam.ferreira.23,hazra.coleman.23}.

In this paper, we discuss the electronic band structure and model parameters, which can be obtained from the {\it ab initio} study.
Recently the SC phase of CeRh$_{2}$As$_{2}$ was analyzed using several theoretical models~\cite{mockli.ramires.21b,mockli.ramires.21,nogaki.daido.21}.
Such models are based on the (intra- and inter-layer) hoppings between the locally noncentrosymmetric Ce atoms.
In this context, it is important to determine an energy scale of the Rashba spin--orbit coupling and the interlayer hoppings~\cite{mockli.ramires.21b}.
The adequate model parameters can be obtained from the the {\it ab inito} calculations combined with the Wannier decomposition of the  electronic band structure.
From the other hand, the quality of the obtained electronic band structure or the Fermi surface renormalization due to the correlation effects~\cite{hafner.khanenko.22} can be examined by the comparison with recently performed the angle-resolved photoemission spectroscopy (APRES) studies of CeRh$_{2}$As$_{2}$~\cite{wu.zhang.23,chen.wang.23}.

The paper is organized as follows. 
First, we provide the details of the computational details (Sec.~\ref{sec.theo}). 
Next, the results obtained are presented in Sec.~\ref{sec.res}. 
In particular, we discuss the basic electronic properties of CeRh$_{2}$As$_{2}$ (Sec.~\ref{sec.el_band}).
We also study the surface states using the the tight binding model in the maximally localized Wannier orbitals, which is based on the direct calculation of the band structure (Sec.~\ref{sec.surface}).
Finally, we discuss the model parameters of our system (Sec.~\ref{sec.model}).
A summary of the results found is presented in Sec.~\ref{sec.sum}.

%%%%%%%%%%%%%%%%%%%%%%%%%%%%%%%%%%%%%%%
%%%%%%%%%%%%%%%%%%%%%%%%%%%%%%%%%%%%%%%
%%%%%%%%%%%%%%%%%%%%%%%%%%%%%%%%%%%%%%%

\begin{figure}[!t]
\centering
\includegraphics[width=0.9\linewidth]{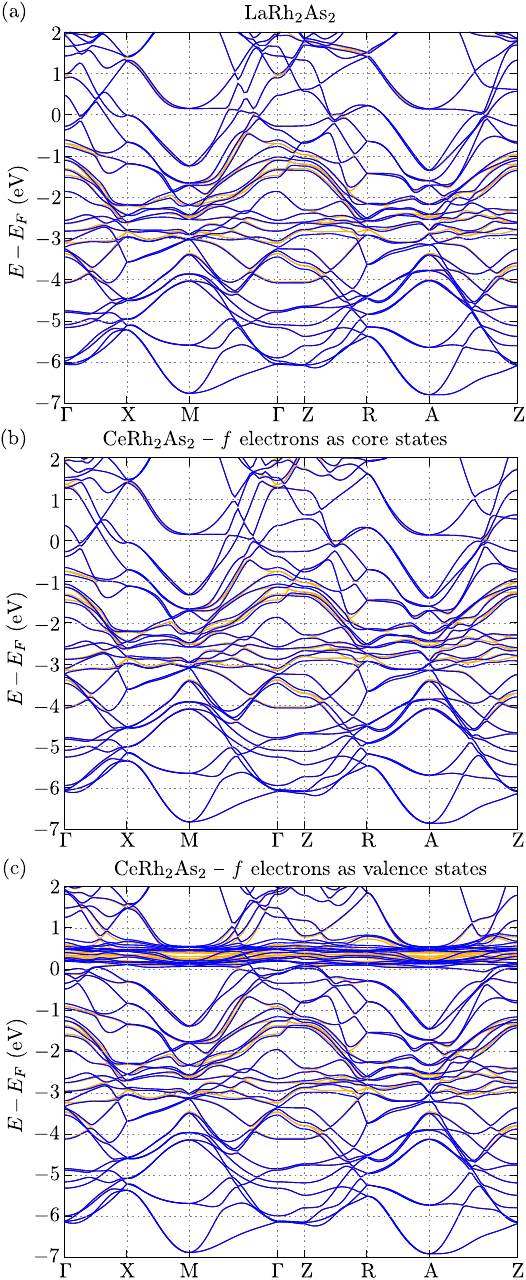}
\caption{
Comparison of the electronic band structures of LaRh$_{2}$As$_{2}$ and CeRh$_{2}$As$_{2}$ (as labeled).
In the case of CeRh$_{2}$As$_{2}$, results for $f$ electrons treated as a core states and valence states are presented (middle and bottom panel, respectively).
\label{fig.bands}
}
\end{figure}

\begin{figure}[t!]
\centering
\includegraphics[width=\linewidth]{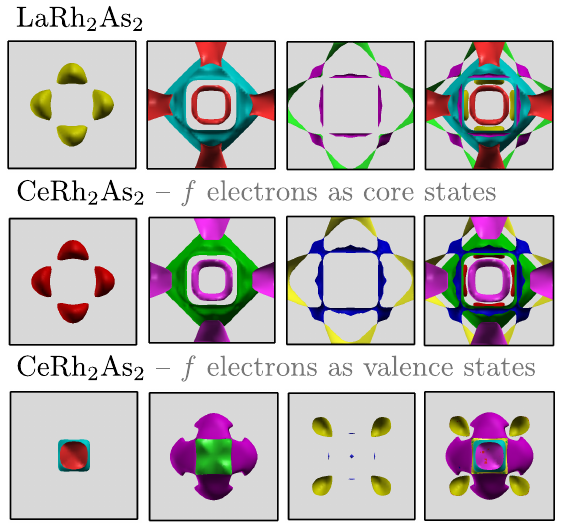}
\caption{
Comparison of the Fermi surface (view for the top) for LaRh$_{2}$As$_{2}$ and CeRh$_{2}$As$_{2}$ (as labeled).
The first two columns present the separate Fermi pockets, while the last column shows the total Fermi surface.
In the case of CeRh$_{2}$As$_{2}$, results for $f$ electrons treated as a core states and valence states are presented (middle and bottom row, respectively).
\label{fig.fermi}
}
\end{figure}

\begin{figure*}
\centering
\includegraphics[width=0.95\linewidth]{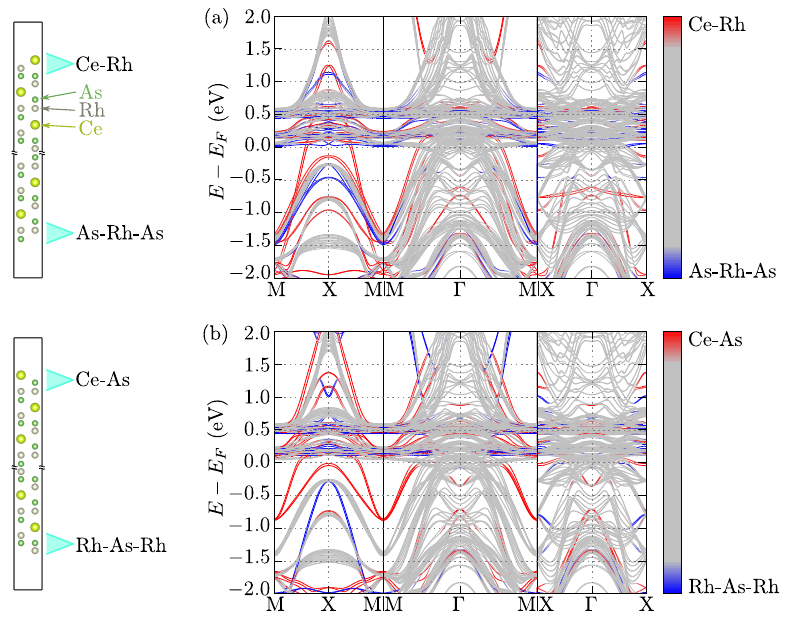}
\caption{
The slab-like calculation for two different surface terminations (crystal structure used is presented on the left side).
The first surface model contains the Ce-Rh and As-Rh-As terminations, while the second surface model includes the Ce-As and Rh-As-Rh terminations.
Right panels present the electronic band structure, while color lines (red and blue) correspond to the projected band structure on the top and bottom surfaces (as presented on the left side).
\label{fig.slab}
}
\end{figure*}

\section{Computational details}
\label{sec.theo}

The first-principles (DFT) calculations are performed using the projector augmented-wave (PAW) potentials~\cite{blochl.94} implemented in the Vienna Ab initio Simulation Package (VASP) code~\cite{kresse.hafner.94,kresse.furthmuller.96,kresse.joubert.99}.
The calculations are made within the revised Perdew--Burke--Ernzerhof for solids (PBEsol) parametrization~\cite{perdew.ruzsinszky.08}, which better reproduce the experimental lattice constant~\cite{ptok.kapcia.21}.
The energy cut-off of the plane-wave expansion is set as $350$~eV. 
The summation over the reciprocal space is performed with a $16 \times 16 \times 8$ \mbox{$\Gamma$-centered} {\bf k}-grid in the Monkhorst--Pack scheme~\cite{monkhorst.pack.76}. 
The calculation includes the effects of the spin-orbit interaction. 
In our calculations, we use the experimental values for the lattice constants, while the atoms positions are optimized.
As a break of the optimization loop, we set the energy differences to $10^{-6}$~eV and $10^{-8}$~eV for ionic and electronic degrees of freedom, respectively.

The surface states are studied by using the tight binding model in the maximally localized Wannier orbitals basis~\cite{marzari.vanderbilt.97,souza.marzari.01,marzari.mostofi.12}. 
This model is constructed from the exact DFT calculations in a primitive unit cell, with $16 \times 16 \times 8$ {\bf k}--point grid, using the {\sc Wannier90} software~\cite{pizzi.vitale.20}.
In the case of $f$ electrons treated as a core states, we construct tight binding model based on La/Ce and Rh $d$ orbitals, and As $p$ orbitals, what corresponds to $42$ orbitals and $84$ bands in the model.
Similarly, for calculation with the $f$ electrons treated as a valence states, we construct model based on La/Ce $d$ and $f$ orbitals, Rh $d$ orbitals, and As $p$ orbitals, what corresponds to $56$ orbitals and $112$ bands in the model.
The electronic surface states are calculated using the surface Green's function technique for a semi-infinite system~\cite{sancho.sancho.85} implemented in {\sc WannierTools}~\cite{wu.zhang.18}.

\begin{figure*}
\centering
\includegraphics[width=\linewidth]{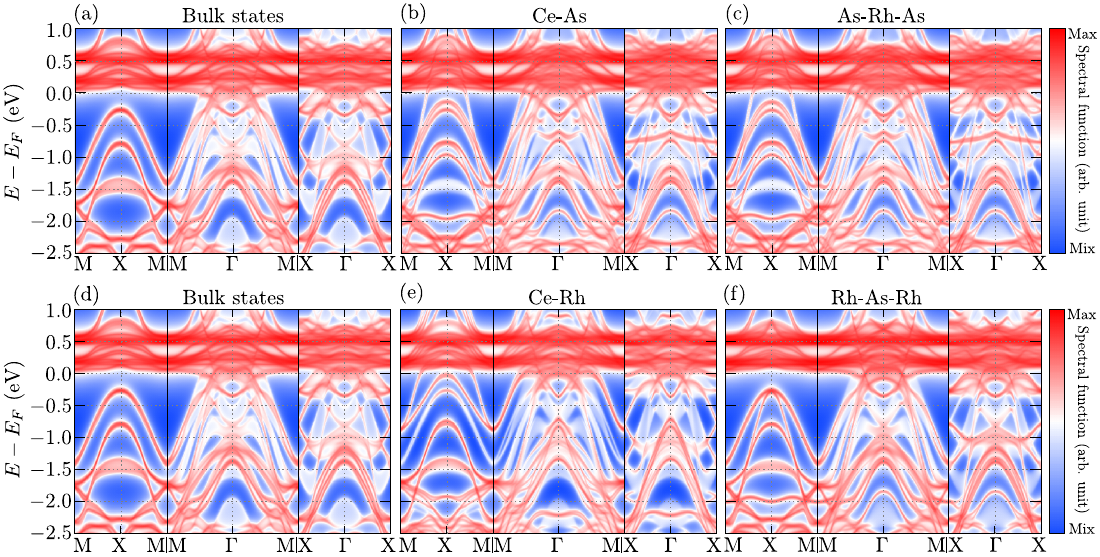}
\caption{
Theoretically obtained spectral functions along the high-symmetry directions for the two-surface models presented in Fig.~\ref{fig.slab} (top and bottom panels, respectively).
Columns from left to right correspond to bulk states, the top surface states, and the bottom surface states (as labeled).
\label{fig.surf}
}
\end{figure*}

%%%%%%%%%%%%%%%%%%%%%%%%%%%%%%%%%%%%%%%
%%%%%%%%%%%%%%%%%%%%%%%%%%%%%%%%%%%%%%%
%%%%%%%%%%%%%%%%%%%%%%%%%%%%%%%%%%%%%%%

\section{Results}
\label{sec.res}

CeRh$_{2}$As$_{2}$ compound crystallizes in the tetragonal structure with the P4/nmm symmetry (space group No.~129).
The experimental lattice parameters are $a = b = 4.283$~\AA, and $c = 9.865$~\AA~\cite{kimura.sichelschmidt.21}.
The atoms are located in five non-equivalent Wyckoff positions.
After optimization the atoms positions were estimated as: Ce $2c$ (1/4,1/4,0.7529), As $2a$ (3/4,1/4,0) and $2c$ (1/4,1/4,0.3635), while for Rh $2b$ (3/4,1/4,1/2) and $2c$ (1/4,1/4,0.1195).
The free parameters (i.e., $z$-components) of atoms positions are close to this reported experimentally~\cite{kim.landaeta.21}.

CeRh$_{2}$As$_{2}$ can be compared with similar material LaRh$_{2}$As$_{2}$, which exhibits the same crystal symmetry.
In this case, the experimental lattice constants are $a = b = 4.314$~\AA, and $c = 9.880$~\AA\cite{kimura.sichelschmidt.21}.
Similarly to the previous material, after the optimization we found atoms positions as La $2c$ (1/4,1/4,0.7548), As $2a$ (3/4,1/4,0) and $2c$ (1/4,1/4,0.3644), while for Rh $2b$ (3/4,1/4,1/2) and $2c$ (1/4,1/4,0.1190), what is comparable with the experimental values~\cite{kimura.sichelschmidt.21}.
As one can see, the lattice parameters are mostly the same for both materials.
Thus, LaRh$_{2}$As$_{2}$ can be treated as a reference system without $f$ electrons around the Fermi level, what can be useful for the electronic band structure comparison and analysis.

\subsection{Electronic band structure}
\label{sec.el_band}

The electronic properties strongly depend on the system parameters (i.e., atom positions and lattice constants).
Comparison between results for LaRh$_{2}$As$_{2}$ and CeRh$_{2}$As$_{2}$ is presented in Fig.~\ref{fig.bands}.
In the case of CeRh$_{2}$As$_{2}$, we present results for $f$ electrons treated as core states [Fig.~\ref{fig.bands}(b)] and as valence states [Fig.~\ref{fig.bands}(c)].
In the absence of the $f$ electrons, the electronic band structures of both compound are very similar [cf.~Fig.~\ref{fig.bands}(a) and Fig.~\ref{fig.bands}(b)].
The $f$ electrons for are localized nearly the Fermi level [Fig.~\ref{fig.bands}(c)].
After the introduction of $f$ electrons as valence ones, an additional shift of the Fermi level is also observed~\cite{ptok.kapcia.21}.
These findings lead to relativity complex electronic band structure around the Fermi level, which can be simply tuned by the external magnetic field.

Mentioned sensitivity of the electronic band structure on the Fermi level is well reflected in the shape the Fermi surface (Fig.~\ref{fig.fermi}).
In both cases of LaRh$_{2}$As$_{2}$ and CeRh$_{2}$As$_{2}$ with $f$ electrons treated as a core state, the Fermi surfaces are very similar.
Contrary to this, the Fermi surface for CeRh$_{2}$As$_{2}$ with $f$ electrons treated as a valence states is significantly different , due to the hybridization of the $d$ and $f$ orbitals~\cite{ptok.kapcia.21},
Additionally, any shift of the Fermi level (e.g., by doping or external magnetic field) can lead to the strong modification of the Fermi level -- this effect should be strongest in the case of the 3rd Fermi pocket (see Fig.~\ref{fig.fermi} -- 3rd panel from the left at the bottom row).
Nevertheless, realized ``true'' Fermi surface in CeRh$_{2}$As$_{2}$ occurring in the nature should be reflected in the ARPES results.

Here, we would like point that the Ce $f$ states are only partially occupied.
Bigger part of ``flat'' electronic bands with dominant $f$-orbital components is unoccupied.
Additional effects, introduced the strong correlation on $f$ states, typically leads to strong modification of these bands.
Indeed, such effect is well visible, for example, within theoretical calculation based on the DFT+DMFT scheme~\cite{li.ma.23}.

\begin{figure*}
\centering
\includegraphics[width=\linewidth]{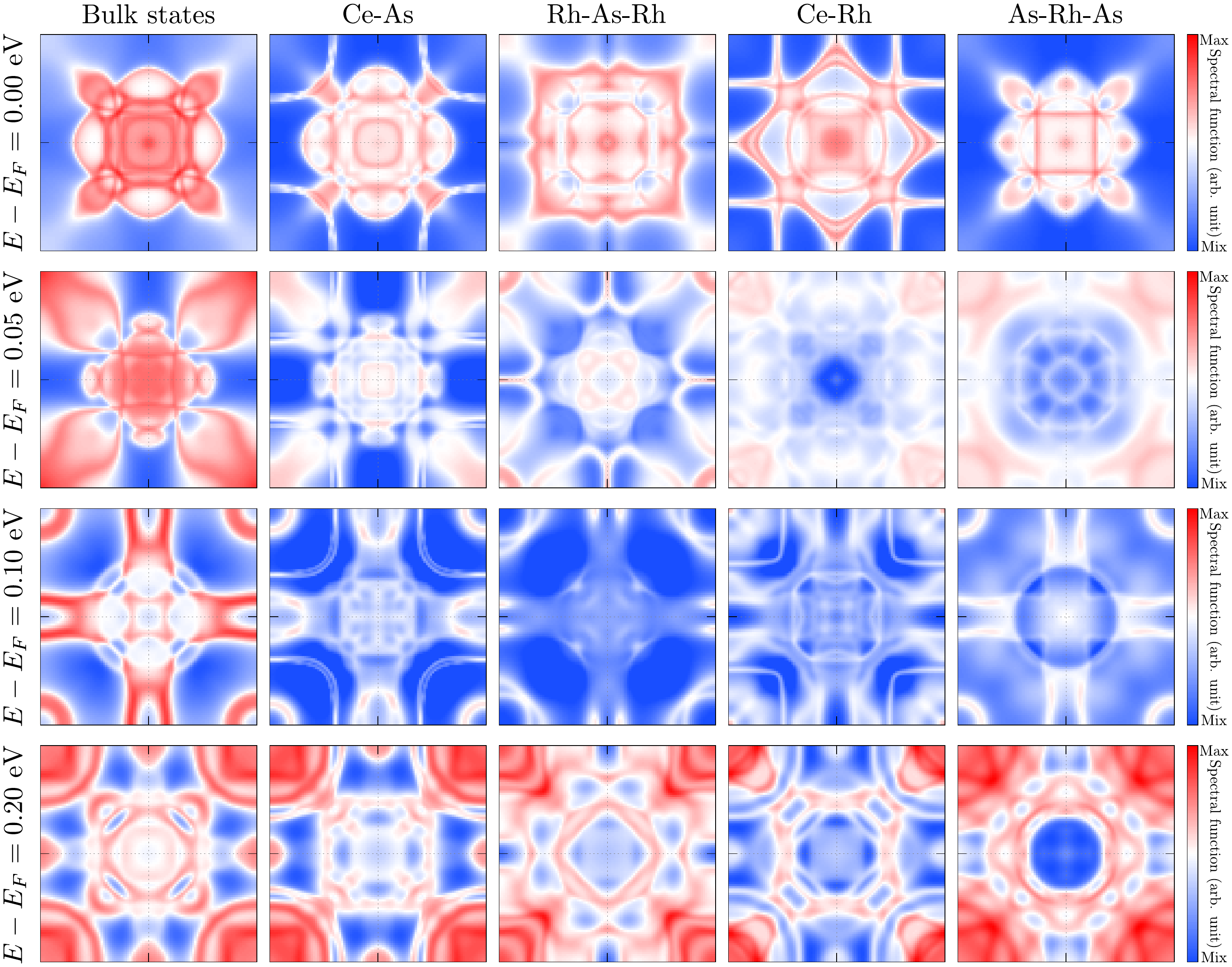}
\caption{
The Fermi surface and constant energy contours (from the top to the bottom, for energies above the Fermi level as given).
Panels from left to right present results for bulk states or surface states with specified surface termination (as labeled).
\label{fig.contour}
}
\end{figure*}

\subsection{Surface states}
\label{sec.surface}

The analyses of the surface states stats from the slab-like calculations (Fig.~\ref{fig.slab}).
A realization of two layers (As-Rh-As and Rh-As-Rh) separated by Ce atoms leads to four possible terminations of the surface (left panels in Fig.~\ref{fig.slab}).
The electronic surface states realized on the top and the bottom surfaces (of the slab) are marked by red and blue color in Fig.~\ref{fig.slab}, while bulk states are presented by grey lines.
As one can expect, the surface termination strongly affect on the occurring surface states.
%%%%%%%%%%%%%%% red
In the case of surface terminated by Ce atoms (red lines), the well-visible surface states are realized along M-X-M path. 
Also along M-$\Gamma$-M path some states are visible.
However, in the case of Ce-Rh termination [the top surface in Fig.~\ref{fig.slab}(a)], the surface states can be realized very closed to the bulk states.
%%%%% blue
In the case of the surface terminated by As-Rh-As layer, the surface states are mostly realized above the Fermi level.
This surface states are also visible along M-X-M path [blue lines in Fig.~\ref{fig.slab}(a)], but close to the bulk states.
The situation looks similar in the case of Rh-As-Rh layer, but in this case, the surface states along M-X-M are well-separated from the bulk states.

The difference between realized surface states is better visible in direct calculations of the surface Green's functions (Fig.~\ref{fig.surf}). 
For simplicity and readability,  we also present the spectral functions of the bulk states (the left column in Fig.~\ref{fig.surf}), which for both models of the surfaces (schemata in Fig.~\ref{fig.slab}) are the same.
This is obvious and clear, because the bulk states do not depends by the surface termination. 
For all surface terminations (cf. the middle and  the right column in Fig.~\ref{fig.surf}), the surface states can be recognized as a separate line along M-X-M path. 
Contrary to this, along M-$\Gamma$-M and X-$\Gamma$-X path, the spectrum is much more complex.
Again, the $f$ electron states are well visible above the Fermi level in the form of flat bands.

In practice, the ARPES spectra are well reproduced by our calculation.
In the case of the ARPES presented in Ref.~\cite{chen.wang.23}, we should focused on the $\Gamma$-M direction (see Fig.~2 therein).
Around $\Gamma$ point several hole-like bands are observed.
From M to $\Gamma$ point, the mostly-linear band are visible. 
Such bands cross the Fermi level.
An absence of electron-like surface states around $-0.8$~eV excludes the Ce-Rh termination in this case.
Also saddle point at X point is well visible, when we compare slope of the bands along X-$\Gamma$ and X-M directions.
This structure is also observed in the ARPES data (Fig.~3 in Ref.~\cite{chen.wang.23}).
Contrary to the results presented in Ref.~\cite{chen.wang.23}, we do not observed gap at X point near the Fermi level (see Fig.~\ref{fig.bands}).
Realization of such gap should not depend only on the correlations or the hybridization effect, but also on the relative distance between the energy levels of the orbitals.
%%%%%%%
Similarly, the results of this work quite well agree with the ARPES experimental results presented in Ref.~\cite{wu.zhang.23}.
In this case, linear-like bands are well visible, however, the $f$-electron flat band is located below the Fermi level (around $-0.3$~eV), what suggest strong overdoping of the system in this case.
Nevertheless, the band crossing observed experimentally around $-1$~eV is also reproduced (see Fig.~3 in Ref.~\cite{wu.zhang.23}).

Experimentally observed ARPES spectra~\cite{wu.zhang.23,chen.wang.23}, suggest that the experimentally observed Fermi level should be shifted with respect to its location obtained theoretically.
Thus, the Fermi surface and constant energy contours for several energies above the Fermi level are presented on Fig.~\ref{fig.contour}.
The Fermi surfaces consisting of only the bulk states are presented in the first column of the figure.
In the case of the theoretically obtained Fermi level, the Fermi surface from bulk states corresponds directly to those presented in Fig.~\ref{fig.fermi}.
However, the existence of the surface states crossing the Fermi level leads to the strong modification of the predicted Fermi surface and constant energy contours for the slab-like systems (cf.~the left column with other columns in Fig.~\ref{fig.contour}).
As we mentioned in the previous paragraph, even a small shift of the Fermi level leads to the strong modification of the results due to the complex band structure (above the Fermi level) found.
Shifting the Fermi level to highest energies, the states around $\bar{\text{M}}$ point, i.e., the around corner of the Brillouin zone, start to play stronger role (see from top to bottom rows on Fig.~\ref{fig.contour}).
%%%%%%% ARPES FS
Moreover, the cross-like constant energy contour presented in Ref.~\cite{wu.zhang.23} (Fig.~6 therein), suggest shifting of the Fermi level founded here by around $0.1$~eV to higher energies.
Nevertheless, a small difference between the constant energy contour observed experimentally and that presented in Fig.~\ref{fig.contour}, suggests smaller role of correlation on the electronic band structure, which is reflected in the absence of the Fermi surface renormalization (e.g., discussed in Ref.~\cite{hafner.khanenko.22}).

\subsection{Model parameters analyses}
\label{sec.model}

The obtained tight binding model in the maximally localized Wannier orbital basis allow to analyze several system properties.
First, the Wannier orbitals onsite energies give information about the band splitting introduced by the crystal electric field (CEF).
Second, the absolute value of the hopping integral between the Wannier orbitals give information about energy scales in the system.
This information can be useful in the formulation of a ``simple'' tight binding model necessary for a realistic description of CeRh$_{2}$As$_{2}$ system.

\begin{figure}[!t]
\centering
\includegraphics[width=\linewidth]{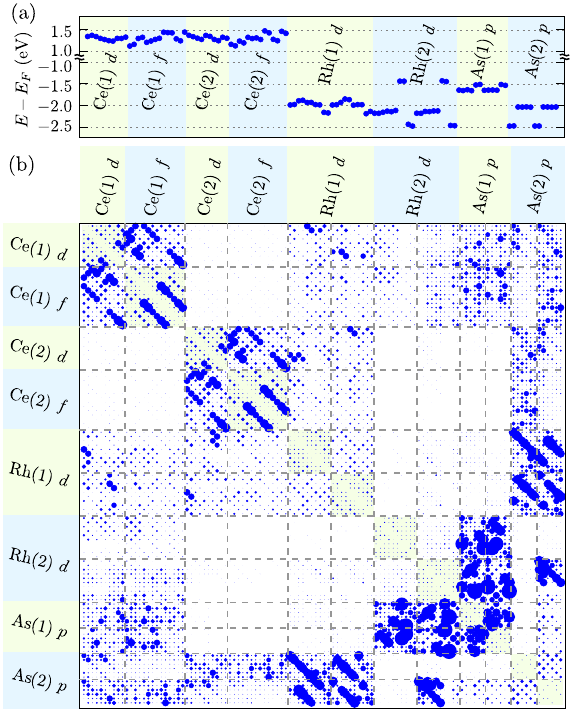}
\caption{
Comparison of (a) the Wannier orbitals onsite energies and (b) absolute values of the hopping integrals between Wannier orbitals (represented by the dot sizes).
\label{fig.onsite}
}
\end{figure}

The model parameters are presented schematically in Fig.~\ref{fig.onsite}.
Here, we introduced a distinction between atoms depending on their location in the crystal.
For the structure presented in Fig~\ref{fig.schem}(a), we get: Ce(1)-Rh(2)-As(1)-Rh(2)-Ce(2)-As(2)-Rh(1)-As(2)-Ce(1) going from the top to the bottom.
As one can see, As(1) and Rh(1) atoms form the square lattice and they are located between Rh(2) and As(2) atoms, respectively.

Let us start the discussion from the onsite energies [Fig.~\ref{fig.onsite}(a)].
In the case of Ce atoms, the $d$ and $f$ orbitals have comparable onsite energy of $1.32$~eV above the Fermi level.
The CEF mostly does not depend on the position and is around $1$--$50$~meV for $d$ orbitals or $10$--$145$~meV for $f$ orbitals.
The spin--orbit interaction introduces additional onsite energy splitting in range of $10$--$20$~meV for $d$ orbitals, and $6$--$28$~meV for $f$ orbitals.
%%%%%
Next, the $d$ orbitals of Rh atoms are located around energy $-2$~eV.
Difference between onsite energy of Rh(1) and Rh(2) orbitals is around $80$~meV.
In the case of Rh(1) atoms, the CEF is around $9$--$165$~meV, while the SOC splitting is around $1$--$40$~meV.
Contrary to this, for Rh(2) atoms, the CEF is around $58$--$630$~meV, whereas the SOC splitting is found around $2$--$5$~meV.
%%%%%
Finally, the $p$ orbitals of As atoms are located around energy $-1.6$~eV for As(1) and $-2.18$~eV for As(2).
In this case, the difference for the onsite energies for As atoms is the biggest and shows how important theirs environments are. 
What is surprising, the CEF for both As atoms are comparable, within range $15$--$45$~meV.
The SOC splitting is much stronger in the case of As(1) atoms (around $30$~meV), while for As(2) atoms is mostly unnoticeable (smaller than $2$~meV).
%%%%%%%%%%%%%%%%%%%%%%%%%
Estimated here CEF is much larger than suggester earlier~\cite{hafner.khanenko.22}, what indicates higher Kondo temperature.

The absolute values of the hopping integrals between the Wannier orbitals are presented in Fig.~\ref{fig.onsite}(b).
Every dot corresponds to the hopping integrals between specified orbitals, while a size of dot corresponds to the magnitude of the hopping.
The parameters are divided into blocks describing hoppings (given as overlapping) between specified atomic orbitals (as labeled from the left and from the top in the figure).
The diagonal block (with the yellow background) corresponds to hoppings between orbitals of the same type of atoms (typically intraorbital hoppings between different orbitals).
Similarly, non-diagonal blocks correspond to the hopping between orbitals centered on different atoms.
%%%%%%%%%%%%%%%%%%%%%%%%%%%%%%%%%%%%%%%%
What is interesting, the interatomic hopping between some atoms (e.g., As(1) and Rh(2) or As(1) within As square net) are bigger than the intraorbital hoppings.
In the diagonal blocks, only hoppings between some $d$ orbitals or some $f$ orbitals of Ce are noticeably big (e.g., first four diagonal blocks).
%%%%%%% squares
It is noticeable that the hoppings between the orbitals of atoms creating RhAs-layers are dominating in the system [i.e. Rh(1) and As(2) or As(1) and Rh(2) atoms].
Similarly, inter-atomic hoppings within the square nets are noticeably large [hoppings between Rh(1) orbitals or As(1) orbitals].
%%%%%%% Ce & rest
What is more, the hoppings between Ce(1) and Ce(2) atoms are negligible small, what can be related to the relatively large distance between these atoms or screening of the atoms by the RhAs-layers.
Moreover, one should notice that the hoppings between Ce atoms and other atoms exhibit strong dependence on the relative position. 
For Ce(1) atoms, the hoppings to all Rh and As atoms are significant, while for Ce(2) atoms only hopping to Rh(1) and As(2) atoms is noticeably large. 
In practice, hoppings between from Ce(2) to As(1) and Rh(2) are smaller than $1$~meV.

\begin{figure}[!t]
\centering
\includegraphics[width=\linewidth]{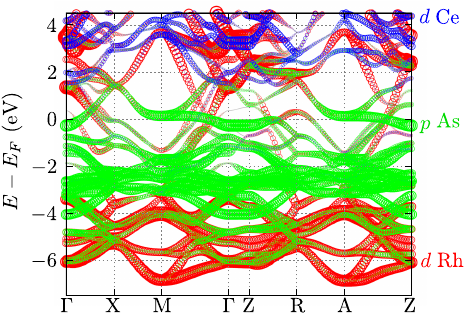}
\caption{
The electronic band structure of CeRh$_{2}$As$_{2}$ projected on selected orbitals.
The results for $f$ states treated as a core states.
Colors of the dots correspond to types of the orbitals, while size of the dots are associated with their contribution strength.
Red, green, and blue colors correspond to $d$ Rh, $p$ As, and $d$ Ce orbitals, respectively.
\label{fig.orbitals}
}
\end{figure}

\section{Summary}
\label{sec.sum}

In this paper, we analyzed the electronic band structure of CeRh$_{2}$As$_{2}$ compound.
We compared the results for this materials with findings for LaRh$_{2}$As$_{2}$.
In the case of the Ce $f$ electrons treated as a core states, the electronic band structures are similar for both materials. 
The Fermi surface also reflect this similarity.
Contrary to this, for the Ce $f$ electrons treated as a valence states, the electronic band structure is more complex above the Fermi surface.
The $f$ states are well visible in a form of nearly-flat bands well-hybridized with other states~\cite{ptok.kapcia.21}.
Such complex structure and close vicinity of the Ce $f$ states to the Fermi level have important role on the obtained results.

Strong sensitivity of the electronic band structure is well visible in the recently presented ARPES spectra~\cite{wu.zhang.23,chen.wang.23}.
In one of them, the Ce $f$ flat states are visible below the Fermi level.
Nevertheless, our study of the slab-like calculation within the surface Green's function method reproduces the observed experimental spectra well.
Depending on the surface termination, the system can realized the surface states visible in the spectral function, what can give information about termination realized experimentally.
Our results clearly reproduce the experimentally observed Fermi surface.
This can suggest a smaller role of the correlation on the electronic band structure then expected.

The recently developed model of CeRh$_{2}$As$_{2}$ is formulated from the Ce-sites perspective~\cite{mockli.ramires.21,mockli.ramires.21b}.
However, performing analyses of the tight binding models based on the maximally localized Wannier orbitals from direct electronic band structure calculations give several interesting information about CeRh$_{2}$As$_{2}$ electronic energy scales.
First, the observed crystal fields are bigger then reported previously~\cite{hafner.khanenko.22}.
Second, the Ce atoms play nonequivalent role in the energetic scale of the system.
Here, the hopping between Ce atoms is negligible small with respect to other hoppings.
Our analyses uncover strong hybridization within the Rh-As layers. 
This opens a new question about more adequate models of CeRh$_{2}$As$_{2}$ describing also the role of As-Rh layers.
%%%%%%%%%%%%
Finally, we would like point that the CeRh$_{2}$As$_{2}$ exhibits several similar properties like simple iron-based superconductor FeSe~\cite{cavanagh.shishidou.22} (both with P4/nmm symmetry).
In the case of iron-based superconductors, the iron-arsenide or iron-selenide layers play important role on the physical properties.
Similarly, for CeRh$_{2}$As$_{2}$, it can be atomic layers formed by As and Rh atoms.
This can be also supported by the exact analyses of the electronic band structure obtained from DFT calculations~\cite{ptok.kapcia.21}.
In the case of CeRh$_{2}$As$_{2}$, the bands crossing the Fermi level are mostly related with As $p$ and Rh $d$ orbitals (see Fig.~\ref{fig.orbitals}).
As a result, the role of the As-Rh layer (and square-like lattice of As and Rh atoms) can play more important role on the physical properties of CeRh$_{2}$As$_{2}$ then initially assumed.
The properties of these layers should be more carefully studied in the future theoretical and experimental investigations.

\begin{acknowledgments}
We kindly thank Daniel F. Agterberg and Aline Ramires for helpful comments and insightful discussions.
Some figures in this work were rendered using {\sc Vesta}~\cite{momma.izumi.11} and {\sc XCrySDen}~\cite{kokalj.99} software. 
A.P. is grateful to Laboratoire de Physique des Solides in Orsay (CNRS, University Paris Saclay) for hospitality during a part of the work on this project. 
A.P. kindly acknowledges the support by the National Science Centre (NCN, Poland) under Project No.~2021/43/B/ST3/02166.
\end{acknowledgments}

%\nocite{*}
\bibliography{biblio}

%%%%%%%%%%%%%%%%%%%%%%%%%%%%%%%%%%%
%%%%%%%%%%%%%%%%%%%%%%%%%%%%%%%%%%%
%%%%%%%%%%%%%%%%%%%%%%%%%%%%%%%%%%%

\end{document}